 \documentclass[aps,prb,twocolumn,twoside]{revtex4}   

\usepackage[dvips]{graphics,graphicx}

\newcommand{\araa} {Ann.\ Rev.\ Astron.\&\ Astrophys.}
\newcommand{\aapr} {Astron.\&\ Astrophys.\ Rev.}
\newcommand{\aap}  {Astron.\ \& Astrophys.}

\newcommand{\apss} {Astrophys.\ \& Space Sci.}
\newcommand{\apjs} {Astrophys.\ J.\ Supp.}

\newcommand{\pasp} {Pub.\ Astron.\ Soc.\ Pac.}

\newcommand{\apjl} {Astrophys.\ J.\ Let.}
\newcommand{\physscr} {Physica Scripta}
\newcommand{\mnras} {Mon.\ Not.\ Roy.\ Astron.\ Soc.}

\newcounter{ion}
\newcommand{\eli}[2]{\setcounter{ion}{#2}#1{~\sc\roman{ion}}}
\newcommand{\mone}{^{-1}}

\newcommand{\pthree}{^3}

\newcommand{\Rev}[1]{{\bf #1}}
\renewcommand{\Rev}[1]{#1}  
\usepackage{changebar}

\begin{document}

\nochangebars

\title{Chandra X-ray Observatory high-resolution x-ray spectroscopy of
  stars: modeling and interpretation}

\author{David P. Huenemoerder}
\email[]{dph@space.mit.edu}


\affiliation{Massachusetts Institute of Technology}

\date{\today}

\begin{abstract}
  \cbstart The Chandra X-ray Observatory grating spectrometers allow
  study of stellar spectra at resolutions on the order of 1000. Prior
  x-ray observatories' low resolution data have shown that nearly all
  classes of stars emit x-rays. Chandra reveals details of line and
  continuum contributions to the spectra which can be interpreted
  through application of plasma models based on atomic databases.  For
  cool stars with hot coronae interpreted in the Solar paradigm,
  assumption of collisional ionization equilibrium allows derivation
  of temperature distributions and elemental abundances. Densities can
  be derived from He-like ion's metastable transition lines.  Abundance
  trends are unlike the Sun, as are the very hot temperature
  distributions. For young stars, there is evidence of accretion
  driven x-ray emission, rather than magnetically confined plasma
  emission. For some hot stars, the expected emission mechanism of
  shocked winds has been challenged; there is now evidence for
  magnetically confined thermal plasmas. The helium-like line emission
  in hot stars is susceptible to photoexcitation, which can also be
  exploited to derive wind structure.  \cbend
\end{abstract}


\maketitle

\section{Introduction}
X-ray emission is ubiquitous among late-type and pre-main sequence
stars, as has been amply demonstrated by imaging and
low-resolution x-ray observatories.\citep{Feigelson:Montmerle:1999}
With the advent of the Chandra transmission-grating and the XMM-Newton
reflection-grating spectrometers, we can now probe the nature of the
x-ray emission in detail through high-resolution diagnostics.  Early
Chandra results confirmed some of the abundance anomalies derived from
low-resolution imaging spectra and also unambiguously confirmed that
few-component temperature models generally are not adequate.  Chandra
spectroscopy has also challenged long-standing hot-star wind and x-ray
production theories. Much effort is now being spent to survey and
analyze stellar x-ray spectra over a range of evolutionary states,
spectral types, rotational periods, and activity levels.

Here we will examine some results for a variety of stars, the
``active'' binaries; young, low-mass stars; and hot, high mass stars,
with emphasis on Chandra grating spectrometers.  We will not discuss
results from the XMM-Newton observatory, though they are complementary
in many ways.

\section{A Brief Introduction to Chandra}
The Chandra X-ray Observatory (CXO) was launched in 1999 and is one of
NASA's Great Observatories.  It has arc second scale imaging x-ray
optics, two transmission gratings, and several detector arrays, two
dedicated to dispersive spectroscopy.  The dispersive spectrometers
are complimentary in resolution, wavelength coverage, and sensitivity.
The High Energy Transmission Grating Spectrometer (HETGS) covers the
range from 1--15 \AA\ at $\Delta\lambda\sim0.01$ \AA\ and 1--30 \AA\ 
at $\Delta\lambda\sim0.02$ \AA\ (concurrently). The Low Energy
Transmission Grating Spectrometer (LETGS) covers 1--175 \AA\ with
$\Delta\lambda\sim0.05$\Rev{\AA}.  \citet{Canizares:00} describe the HETGS and
initial observations of Capella, and \citet{Brinkman:2000} do
similarly for LETGS.  \citet{Weisskopf:02} describe all the
instruments and capability of CXO.

\section{Stellar X-ray Sources}
Stars of nearly all spectral types (a surface temperature
classification) are significant x-ray emitters.  The only exceptions
are the A-type stars ($T_\mathrm{eff}\sim10^4\,\mathrm{K}$) whose
atmospheres are too placid to generate x-ray-emitting structures.
Cooler stars are magnetically active. In analogy with the Sun, they
are presumed to have coronal structures driven by a magnetic dynamo,
since there is a strong correlation of x-ray luminosity with
rotation rate.\citep{Pallavicini:1989, Walter:81} These stars also are
known to have dark starspots, strong ultraviolet emission, and
activity cycles.\citep{Baliunas:Nesme:1996,Wilson:1978}  Youthful,
low mass stars are also prodigious sources of x-ray emission.
\citep{Feigelson:Montmerle:1999}  This is thought to be primarily due
to their primordial rapid rotation and associated dynamo, which decays
with age unless there are external driving forces.

Hot, high-mass stars are also strong x-ray sources, but here the
mechanisms are different.  These stars are extremely luminous in the
optical and ultraviolet, and the radiation field can drive a massive
wind, in which instabilities can lead to shocks.
\citep{Seward:1979,Lucy:White:1980,Lamers:Cassinelli:1999}

Degenerate objects (white dwarfs, neutron stars, and black holes) in
close binaries are some of the brightest x-ray emitters, as material
is compressed and heated in unstable accretion disks or on the compact
object's surface.  We will not be discussing these further in this
paper.

\section{Plasma Modeling}
A \Rev{common} assumption of coronal modeling is that the plasma is in
collisional ionization equilibrium (CIE).  This means that ionic
species are \Rev{predominantly} in their ground state, are ionized or
excited by collisions with thermal electrons, and recombine and
radiatively decay. It is also assumed that the plasma is optically
thin, so that appreciable scattering does not occur.  Additionally, it
is usually assumed that the plasma, while it may have temperature and
density gradients, has a uniform \Rev{elemental} composition.

Under CIE, the emitted spectrum is a function of electron temperature,
$T_e$, and elemental abundance, $A_Z$ (for atomic numbers, $Z$).
Since the plasma is highly ionized, most of the electrons come from
hydrogen and helium
\Rev{(astrophysical plasmas are composed primarily of H and He, with a
trace of heavier elements)}.
The spectral energy distribution is mostly only
weakly dependent upon electron density, $n_e$, except for some few
highly density sensitive transitions.

The line luminosity, $L_l$, for spectral feature, $l$, can be expressed
as:
\newcommand{\eqabund}{{A_{Z}}}
\newcommand{\eqdem}{{\left [n_e n_H \frac{dV(T)}{d\log T} \right ]}}
\newcommand{\eqemis}{{\,\epsilon_l(\log T)}}
\begin{equation}
  \label{eq:dem}
       L_l = \eqabund \int{ \eqdem \eqemis\, d\log T}
\end{equation}
Here, $n_H$ is the hydrogen number density, $V$ is the volume of
material at temperature $T$, and $\epsilon$ is the line emissivity as
defined by \citet{Raymond:96} (with units of
$\mathrm{photons\,cm\pthree\,s\mone}$).  The elemental abundances,
$A_Z$, are relative to Solar.

The quantity in square brackets is known as the differential emission
measure ($DEM$).  It is a one-dimensional parameterization of the
emitting volume's temperature structure.  It is, in a sense, the most
one can discern about the structure without additional constraints,
such as obtained by imposition of geometrical and temperature
structure from an ensemble of magnetically confined loops,\citep{RTV}
or by constraining the volume with a spectroscopic density
determination or via some geometric mapping method.  \Rev{
  \citet{Pottasch:1963} presented an early application of Solar
  spectra emission measure modeling.}  \citet{Griffiths:98} gave
detailed application of emission measure modeling to ultraviolet
spectra, and also cite significant historical work.

The functional dependence of $\epsilon$ in Equation~\ref{eq:dem} makes
direct inversion impossible \citep{Craig:76, Hubeny:95} (even {\em
  without} the additional degeneracy imposed by the abundance factor,
$A_Z$).  The $\epsilon$ can be thought of as a set of basis vectors;
in the $\log$, they are \Rev{often} approximately parabolic with full
width-half maximum near 0.3 dex, and they overlap greatly from species to
species, providing more redundant than unique information.  Hence, we
must rely on forward-folding, iterative techniques with {\em a priori}
biases.  Some authors use polynomial parameterizations of the
$DEM$\citep{Schmitt:Ness:2003,Lemen:1989} and others sophisticated
statistical methods which formally incorporate sources of uncertainty.
\citep{Kashyap:98}

We chose a simple, fairly brute-force approach: we minimize
Equation~\ref{eq:dem} for the $DEM$ and abundances on an arbitrary
temperature grid for as many lines as can be reliably identified and
whose flux can be determined from parametric fits to the spectral
features.  Since the fit is underdetermined, we use a regularization
term which imposes some smoothness on the $DEM$ solution.  In order to
estimate the effect of observational statistical uncertainties, we
implement a Monte-Carlo iteration in which the measured line fluxes
are perturbed by their uncertainties, and about 100 iterations are
averaged.  In this way, we simultaneously estimate the $DEM$ and the
abundances.  The solution is not unique, but is consistent and must
adequately predict the spectrum. An example of application to the
active binary, AR~Lacertae, is given by
\citet{Huenemoerder:Canizares:al:2003}  Figure~\ref{fg:impeghegvsmeg}
shows a small portion of the HETGS spectrum of long-period (24 days),
single-lined spectroscopic binary, IM~Pegasi, along with a synthesized
spectrum using the $DEM$ and abundance solution.

\subsection{Atomic Data}
The importance of the atomic database cannot be overestimated.  The
emissivities in Equation~\ref{eq:dem} embody a large community effort.
To identify lines, look up emissivities, and synthesize line and
continuum spectra, we use the Astrophysical Plasma Emission Database
(APED).  \citep{Smith:01} This database includes effective collisional
excitation rate coefficients, photoionization rates, dielectronic
satellite line strengths, and atomic transition probabilities,
spectroscopic designations for many levels, and, as available,
wavelengths accurate enough for high-resolution spectroscopy.
Ionization balance models are also included.  The APED is available as
a set of files in standard portable formats and also via an on-line
interactive browser.
This plasma database 
is also the underpinning of the Interactive Spectral Interpretation
System \citep[ISIS]{Houck:00} which we use for spectral measurement
and modeling.

\subsection{The Importance of Resolution}
\citet{Kastner:02} 
show a dramatic comparison of the previous
low-resolution imaging x-ray spectra compared to the Chandra HETGS.
We can now see individual features that were only inferred before,
sometimes incorrectly: \citet{Kastner:1999} modeled the emission as
due primarily to iron.  The high resolution spectrum showed that iron
was highly depleted in the corona, and that neon was very
strong.\citep{Kastner:02}
  
\begin{figure*}
  \scalebox{1.3}{\includegraphics[angle=-90]{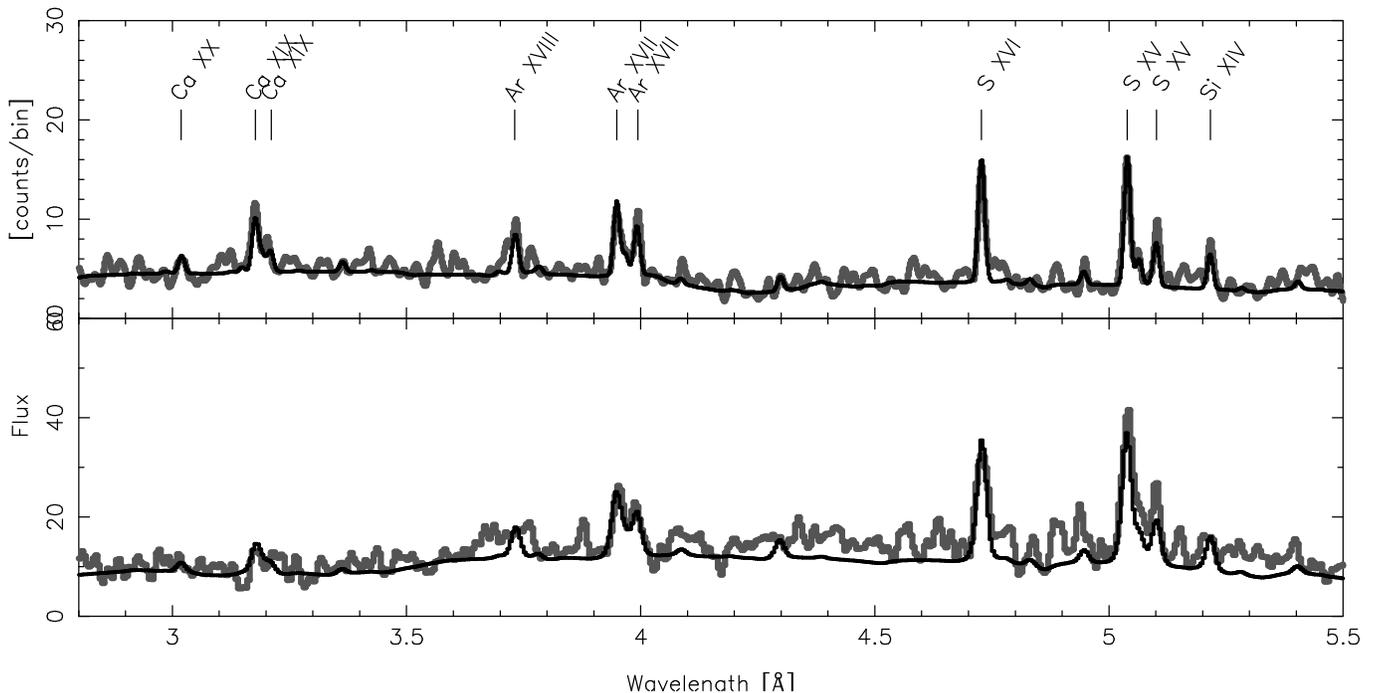}}
  \caption{A portion of the IM~Pegasi HETGS spectrum demonstrates the
  importance of resolution in line detection and measurement.  The top
  panel shows the HEG spectrum, which has twice the resolving power of
  the MEG data (lower panel).  Some emission lines, which are distinct
  in the HEG data, may not even appear as measurable features from the
  MEG. The smoother curve is a model, convolved with the
  instrumental response.  The data are the sum of eight 25 ks
  exposures.
  \label{fg:impeghegvsmeg}}
\end{figure*}

Even given the CXO dispersive spectrometers, resolution can be
crucial, and can outweigh a factor of two difference in overall
sensitivity.  In Figure~\ref{fg:impeghegvsmeg} we show a portion of
the HETGS spectrum of IM~Pegasi, a K1~III, RS~CVn-type binary.  The
HETG is comprised of two grating types, the HEG and MEG (high and
medium energy gratings, respectively).  HEG has about twice the
resolution as MEG at the same wavelength and order, but only half the
effective area.  In the HEG, we can quite clearly see and measure
lines, such as \eli{Ca}{19} $\lambda3.2\,$\AA, \eli{Ar}{18}
$\lambda3.7$ \AA, or \eli{Si}{13} $\lambda5.2\,$\AA, which are not
obvious at MEG resolution.

\section{Results for Active Binaries}

The Solar corona is a common reference for interpretion of active
binary spectra.  Active binary stars are, however, three to four
orders of magnitude more luminous in x-rays than the Sun.
\citep{Walter:81, Dempsey:Linsky:al::1993a}  It was not entirely
surprising, then, when \Rev{these stars showed very non-Solar
  characteristic coronal temperatures and abundances.}

\subsection{Elemental Abundances}
\Rev{High resolution spectroscopy has allowed more detailed evaluation
  of abundance anomalies first derived from low resolution data.
\citep{Singh:96,Huenemoerder:Canizares:al:2003}
}
The Solar first ionization potential (FIP) effect
\citep{Feldman:90,Feldman:92,Feldman:Laming:00} is not followed.
Instead, the trend is the reverse: high FIP elements are enhanced,
though it is not a totally uniform or consistent trend.  Examples may
be found in \citet{Brinkman:01,Audard:01a,Huenemoerder:01}

\begin{figure}
  \scalebox{0.6}{\includegraphics[angle=0]{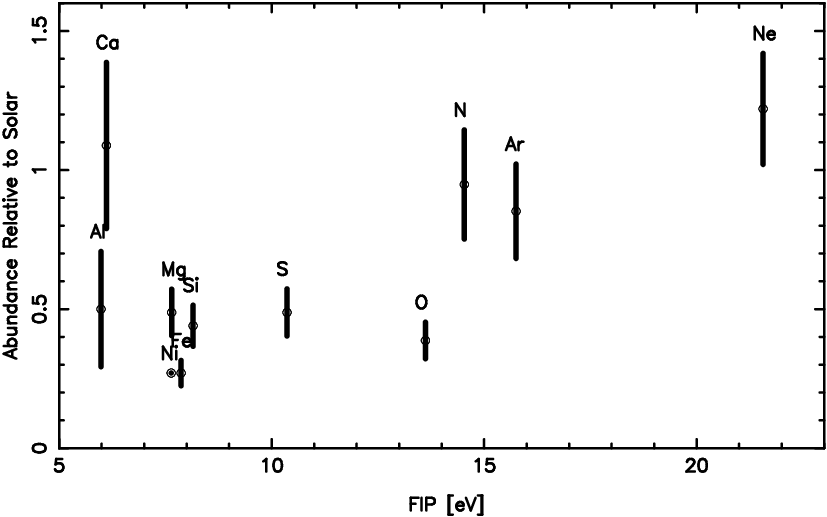}}
  \caption{Abundances relative to Solar photospheric values for the
    IM~Peg corona, as derived from the emission-measure and abundance
    fit to line fluxes.  Uncertainties are the result of a
    Monte-Carlo iteration in which measured line fluxes were perturbed
    by their statistical measurement uncertainty. The $x$-axis is the
    first ionization potential.
\label{fg:impegabund}}
\end{figure}

We have derived a preliminary set of abundances for IM~Pegasi, for the
same observations shown in Figure~\ref{fg:impeghegvsmeg}.  They are
rather typical results for coronally active binaries, with depleted Fe
and enhanced Ne.  Figure~\ref{fg:impegabund} shows the abundances
relative to solar photospheric values versus the first ionization
potential. 

The FIP is a convenient coordinate, but ultimately
may not be physically significant. In the Sun, the coronal abundance
drops by a factor of four at about 10 eV.\cite{Feldman:Laming:00}
\citet{Huenemoerder:Canizares:al:2003} compare the AR~Lac results to
Solar. 

While we generally suspect that the abundance anomalies are important,
we have no physical mechanism for predicting the fractionation, even
in the Solar case.  An additional complication we face with stellar
coronae is that we often do not know the stellar photospheric
abundances; they are difficult to determine from photospheric lines
blended by rapid rotation.

The relative and absolute abundances must ultimately relate to the
various mixing and segregating forces at work: gravitational settling,
thermal diffusion, electrostatic forces, and bulk flows, for examples.
In some cases, it appears that the relative abundances truly reflect
an underlying physical phenomenon.  As stars evolve, CNO processing
occurs in the core, and can be dredged to the surface, and in binary
systems, transferred to a companion in a dynamical mass-transfer
phase. \citet{Drake:03b,Drake:Sarna:2003} explore two scenarios with
LETGS spectroscopy of C and N in a few stars and clearly show strong
differences in the C and N line strengths in otherwise similar stars,
consistent with stellar evolutionary theory.

\subsection{Temperature Structure}
As is obvious from Equation~\ref{eq:dem}, temperature structure (the
$DEM$) and abundances are not determined independently.  They are
highly degenerate if there is no coupling from element to element by
overlapping emissivity functions.  Fortunately, there are many ion
states of iron present, typically from \eli{Fe}{17} to \eli{Fe}{25}.
Many authors first derive the $DEM$ for Fe, then adjust abundances of
other elements to bring the inferred emission measure into agreement
with that of Fe.  This is acceptable if the distributions strongly
overlap.  At the low end of the temperature range available to
Chandra, however, there is little overlap of Fe with O or N, for
example. In this case, we rely on the simultaneous fit to
``bootstrap'' N to O to Fe.  A representative view of the HETGS
line-temperature coverage can be found in
\citet{Huenemoerder:Canizares:al:2003} (which also includes some ions
from the extreme ultraviolet range).

We frequently find that the $DEM$ is double-peaked with components
near $\log T\sim 6.8$ and $\sim7.4$.
\citep{Huenemoerder:Canizares:al:2003,Drake:01}  Weak, but
significant, tails are seen on both the high and low sides.  For the
case of AR~Lac, we had supplemental extreme ultraviolet data to extend
the solution below $\log T=6.5$, and found that the minumum in the
x-ray regime is real.  On the hot end, the fit is poorly constrained
since there are only a few weak lines with broad emissivity functions.
The continuum also provides a high-temperature constraint, but also
with low temperature resolution.  That there is very hot material is
undisputed, but the shape of the hot tail is not unique.

\begin{figure}
  \scalebox{0.6}{\includegraphics[angle=0]{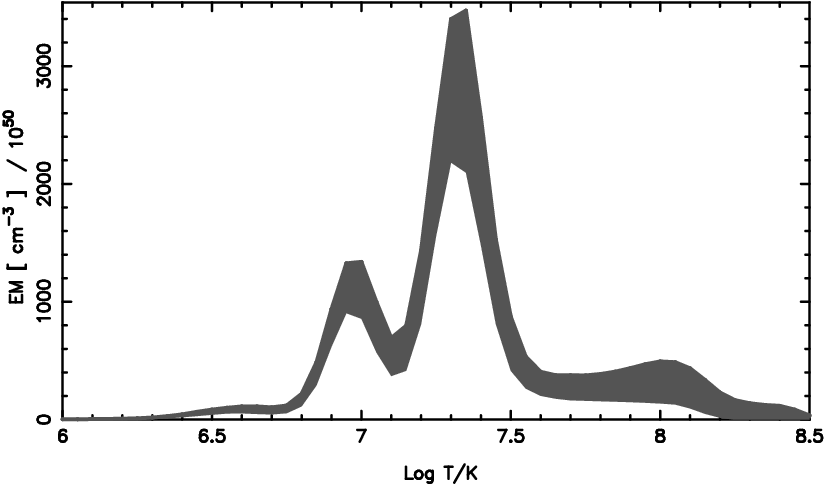}}
  \caption{$DEM$ for IM~Peg, from the same model in
    Figure~\ref{fg:impegabund}. The width of the band is the standard
    deviation from the Monte-Carlo iteration.
\label{fg:impegdem}}
\end{figure}

The solution to IM~Pegasi (the same fit whose abundances are shown in
Figure~\ref{fg:impegabund}) is typical of active binaries and is
shown in Figure~\ref{fg:impegdem}. 

We believe that the hotter peak is primarily due to flares, in analogy
with impulsive magnetic reconnection events on the Sun which heat
plasma.  In the case of II~Pegasi, we had enough data and a
conveniently timed flare to model the $DEM$ in both pre-flare and
flare states, and it was clear that the hot peak was due to the
flare.\citep{Huenemoerder:01}  For AR~Lac, with fewer flare counts, we
relied on a proxy indicator: line flux modulation binned by
temperature of formation, which showed that high $T$ lines --- roughly
coincident with the $DEM$ hotter peak --- were highly modulated, but
cooler lines were not.\citep{Huenemoerder:Canizares:al:2003} \Rev{Capella
has only one peak near $\log T=6.8$, but does have a variable hot
component which could be due to some lower level of flare
contribution.}\cite{Brickhouse:00}

\subsection{Geometric Structure}
The geometry of stellar coronae is of great interest.  In the Sun, we
have several different coronal structures: magnetically confined
loops; open-field regions which merge the corona with the
interplanetary medium; flaring loops; eruptive prominences; and bright
points, to name a few.  For other stars, we can only infer or
indirectly image the surfaces and coronae. One method for determining
scale is to search for rotational or eclipse modulation.
\citet{Brickhouse:01} found modulation in 44 Boo, a short period
eclipsing system, consistent with polar active regions on one
component.  In order to discriminate fluctuations due to flares, the
high-resolution spectrum is necessary in order to exclude features
formed predominantly at flare temperatures.

\begin{figure}
  \scalebox{0.6}{\includegraphics[angle=0]{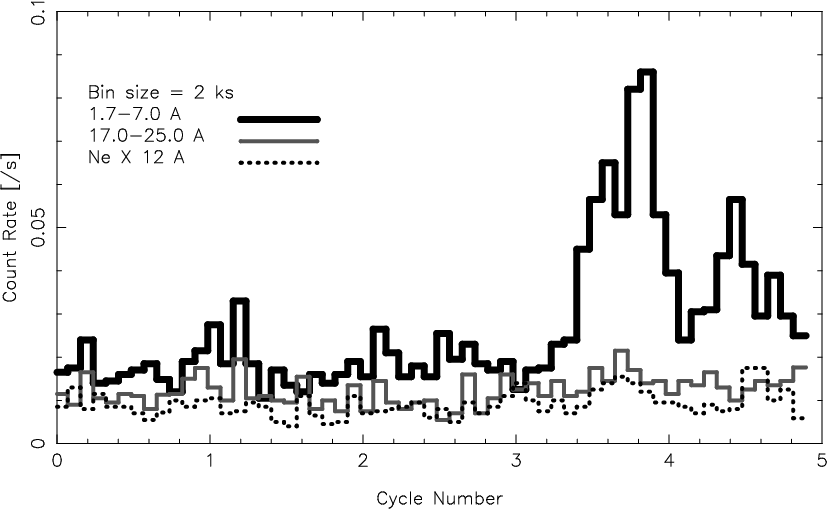}}
  \caption{The counts in 2000 second bins {\it vs.}\ phase cycle are shown
    in 3 bandpasses for VW~Cep, an 0.25 day period coronally active
    binary.  The short-band curve (1.7--7 \AA; upper, thick solid
    line) shows a sequence of large flares beginning near phase 3.5.
    The narrow bands (two lower curves; solid gray is 17--25 \AA,
    and dotted is \eli{Ne}{10} 12 \AA), which are formed predominantly
    at lower temperatures than the short band, do not show the flares.
    \label{fg:vwceplc}}
\end{figure}

This effect is clearly demonstrated in a light curve of VW~Cep, an
0.25 day period binary system. Figure~\ref{fg:vwceplc} shows the count
rate over 5 rotations in a short wavelength band (1.7--7 \AA), the
17--25 \AA\ region (which is dominated by iron and oxygen emission
lines), and in the \eli{Ne}{10} 12.3 \AA\ line (which is blended with
\eli{Fe}{17}).  It is obvious that the large flare is not seen in the
low-temperature features.  There is a hint that \eli{Ne}{10} is
systematically decreased near phase modulus 0.5; phase-folded curves
will reveal whether we have detected compact coronal structure, or
whether emission is uniformly distributed (or of large extent,
relative to stellar radii).

Radial velocity information can provide a general locus for coronal
emission.  \citet{Ayres:01} found that the \eli{Ne}{10} centroid in
HR~1099 followed the K1~IV star, indicating that it is the primary
source of x-ray emission. \citet{Huenemoerder:Canizares:al:2003} used
line width variability to determine that in AR~Lac, both components
were equal contributors to the x-ray flux.

\section{Density Diagnostics}
Young stars are well known as prodigious x-ray
emitters,\citep{Feigelson:Montmerle:1999} and x-ray imaging is good
for identifying young stars obscured by dust at optical wavelengths.
The working hypothesis has been that since young stars are rapidly
rotating, they have a strong magnetic dynamo, and they have
\Rev{activity analogous to that of} the active binaries (which rotate
rapidly due to their binary nature and tidal coupling).

Our first high resolution x-ray spectra of a classical T Tauri star
(CTT; a youthful, low mass star, still actively undergoing accretion),
TW~Hydrae shed doubt on this interpretation.\citep{Kastner:02} Neon is
extremely overabundant, and iron very depleted, similar to, but more
extreme than the typical coronal sources.  Since we don't yet know what
fractionates the x-ray emitting plasma, this is still a curiosity
without theoretical underpinning.  The $DEM$ was very cool ($\log
T\sim6.5$) and narrow.  This is different in general from active
binaries, but not extreme, since it is not much different from
Capella.  

The single, starkly contrasting feature is in the ratios of the
helium-like triplets: both \eli{Ne}{9} and \eli{O}{7} in TW~Hya show
ratios indicative of high density.  Coupled with evidence from
other wavelengths for accretion, \citet{Kastner:02}
argued that the emission mechanism is not strictly ``coronal''
(magnetic loops and a dynamo), but somehow generated in an accretion
funnel.

The sensitivity of the helium-like triplets to density has been known
for a some time.\citep{Gabriel:69,Gabriel:73}  The forbidden line is
metastable; above some critical value, it can become depopulated by
collisions, thereby reducing its flux in favor of the intercombination
line. The ratio of the forbidden line ($f$) to the intercombination
($i$) is density sensitive above some critical density determined by
the nuclear charge.  In the Chandra bandpass, the most prominent
He-like triplets are from \eli{Si}{13}, \eli{Mg}{11}, \eli{Ne}{9},
\eli{O}{7}, \eli{N}{6}, and \eli{C}{5} (the latter two are
LETGS-only).  These span a range of critical densities potentially
useful for stellar plasma diagnostics. \citet{Ness:Brickhouse:al:03}
plot the critical densities of these ions.  (There are other triplets
present, such as \eli{S}{15}, \eli{Ar}{17}, \eli{Ca}{19}, and
\eli{Fe}{25}, but the resolution is not sufficient to resolve
them adequately.)

\begin{figure*}
  \scalebox{1.2}{\includegraphics[angle=0]{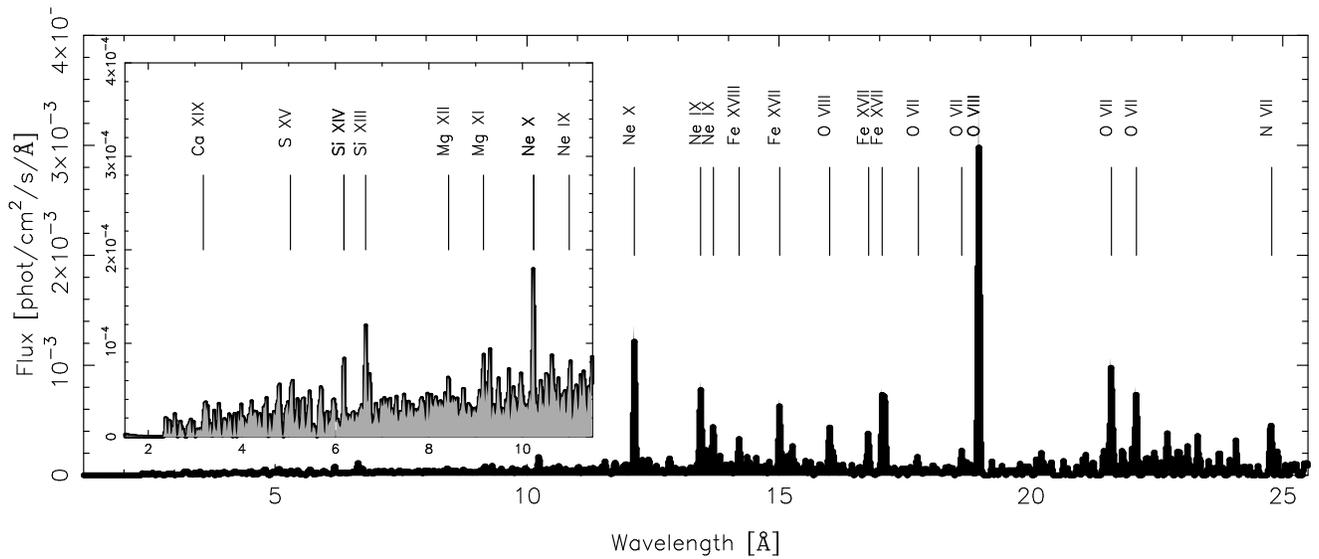}}
  \caption{The Chandra HETGS spectrum of TV~Crt (HD98800), a
    weak-lined T Tauri star.  The inset shows an expanded view of the
    short wavelength region, with significant detections of features
    formed at high temperatures, and which are not present in the
    TW~Hya spectrum.
\label{fg:tvcspec}}
\end{figure*}

\begin{figure}
  \scalebox{0.6}{\includegraphics[angle=0]{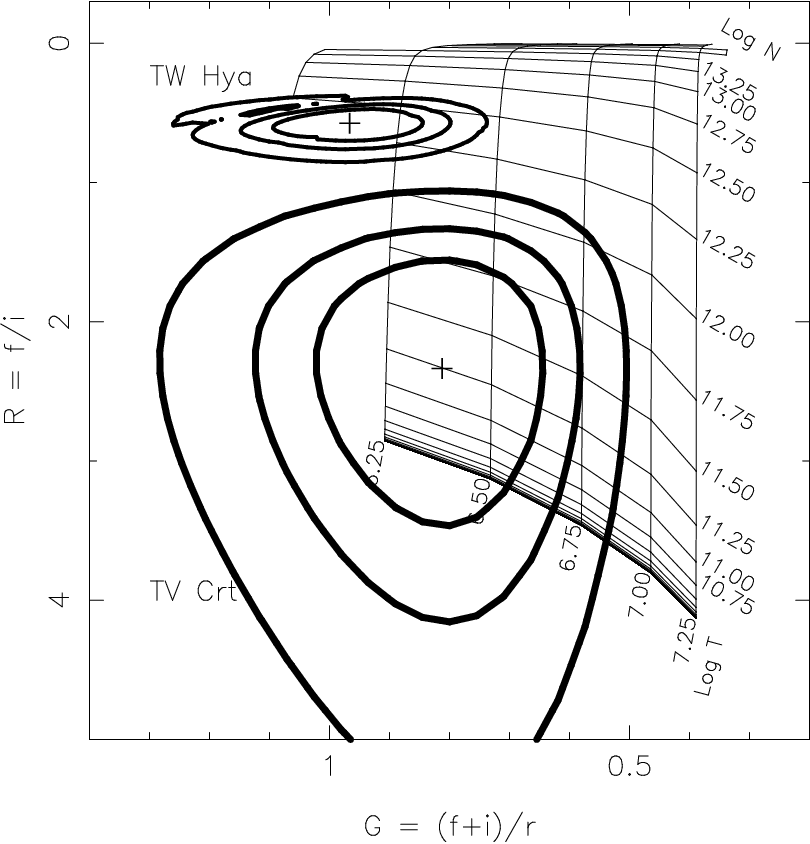}}
  \caption{We show the confidence countours for the \eli{Ne}{9}
    density ($R=f/i$) and temperature ($G=(f+i)/r$) sensitive line
    ratios for TW~Hya and TV~Crt, two pre-main sequence stars in the
    same physical association.  Contours (inner to outer) are the 68,
    90, and 99\% limits.  The small upper oval is for TW~Hya, whose
    density is clearly constrained, and best fit value is about $\log
    n_e=12.5$.  TV~Crt (lower contours) only provides an upper limit
    of 12.25 (99\% confidence).  The grid overlay is from the density
    dependent atomic database, with $\log T$ labeled along the
    approximately horizontal direction, and $\log n_e$ vertically.
\label{fg:tvtwfir}}
\end{figure}

In order to pursue the nature of pre-main-sequence x-ray emission, we
have recently obtained the HETGS spectrum of a weak-lined T Tauri
(WTT) star, TV~Crt (WTTs have presumably finished the accretion stage
of evolution, but have not yet finished gravitational contraction to
the main sequence when nuclear fusion starts and sustains pressure
balance).  We show its spectrum in Figure~\ref{fg:tvcspec}.
The corona appears to be intermediate between TW~Hya and the active
binaries: it has a fairly cool component to the $DEM$, but also a
warmer tail which drops off above $\log
T\sim7$.\citep{Huenemoerder:2004}  To quantify the density diagnostic,
we fit the $f/i$ ratio, and also the ``G''-ratio, $(f+i)/r$ which is
primarily sensitive to temperature ($r$ represents the resonance line
flux).  Figure~\ref{fg:tvtwfir} shows contours of confidence for these
ratios for \eli{Ne}{9} ($\sim13.6$ \AA). The two stars are clearly
distinct at the 99\% confidence level.

\subsection{A Warning}
The positive density detection in TW~Hya was easy, even in a rather
low-signal observation, due to its extreme value.  In other spectra,
even deep observations, the measurement can be a very difficult task.
\citet{Ness:Brickhouse:al:03} present an in-depth study of Capella
using HEG, MEG, LETG, and XMM-Newton grating spectra. They show that
there is essentially no such thing as a model-independent line ratio.
The only way to account for blends is to model the entire spectrum,
then evaluate the relative contributions to any particular spectral
feature using that model.  They applied this to the \eli{Ne}{9}
triplet, and found that about 30\% of the intercombination line is due
to iron.  This 30\% can change the $f/i$ ratio from an apparently
constrained density measurement to unconstrained, even at HEG
resolution.

\section{Hot (High-Mass) Stars}
At the other end of the temperature spectrum are the hot, high-mass
main-sequence stars (O and B spectral types).  The standard model for
x-ray production in these objects was a massive, radiatively driven
wind with shock instabilities.\citep{Lucy:White:1980}  Temperatures
were expected to be low.  Due to outflow and optical depth effects,
line profiles were expected to be blue-shifted and asymmetric.
Chandra/HETGS spectra have changed that view.  Some objects did behave
as expected, such as $\zeta$ Puppis. \citep{Kramer:Cohen:al:2003,
  Gagne:Cohen:al:2001, Cassinelli:Miller:al:2001}  Others did not:
\citet{Schulz:2003} found narrow, symmetric, and un-shifted lines in
$\theta^1$ Ori~C.  It now appears that there can also be magnetically
confined and very hot plasma in hot stars.  There is currently a fairly large
observational and theoretical campaign to explore and explain this
unexpected behavior.  Models of $\theta^1$ Ori~C, which is
periodically variable and has a measured magnetic field, have a wind
channeled into a shock by primordial magnetic
fields.\citep{Gagne:Cohen:al:2001} It remains to be determined
whether the previous x-ray/wind theory is the rule or the exception.

\begin{acknowledgments}
  This work was supported by Smithsonian Astrophysical Observatory
  contract SV3-73016 to MIT for the Chandra X-Ray Center, and by
  Chandra award numbers G02-3005A \& G03-4005A issued by CXC, which is
  operated by SAO for and on behalf of NASA under contract NAS8-39073.
  I also thank various collaborators and kibbitzers: N.~Schulz,
  B.~Boroson, C.~Canizares (MIT); J.~Kastner (RIT); D.~Buzasi,
  H.~Preston (USAFA); J.~Drake, N.~Brickhouse (CfA).
\end{acknowledgments}

\clearpage
\raggedright

\end{document}